\documentclass[twocolumn]{jpsj2}
\title{Interplay between Charge and Magnetic Orderings in YbPd}

\author{Atsushi \textsc{Miyake}$^1$\thanks{E-mail address: miyake@issp.u-tokyo.ac.jp}\thanks{Present address: The Institute for Solid State Physics, The University of Tokyo, Kashiwa, Chiba 277-8581, Japan},
Kazuki \textsc{Kasano}$^1$,
 Tomoko \textsc{Kagayama}$^1$, Katsuya \textsc{Shimizu}$^1$, \\
Ryo \textsc{Takahashi}$^2$, Yusuke \textsc{Wakabayashi}$^2$, Tsuyoshi \textsc{Kimura}$^2$ and Takao \textsc{Ebihara}$^{3}$}

\inst{$^{1}$KYOKUGEN, Osaka University, Toyonaka 560-8531, Japan,\\
 $^{2}$Graduate School of Engineering Science, Osaka University, Toyonaka 560-8531, Japan,\\
 $^{3}$Department of physics, Shizuoka University, Shizuoka 422-8529, Japan}

\abst{The pressure-temperature phase diagram of an intermediate-valence compound YbPd was revealed via simultaneous ac-calorimetry and electrical resistivity measurements.
Two successive structural phase transition temperatures, $T_1$ = 125~K and $T_2$ = 110~K, are found to decrease with increasing pressure.
The magnetic ordering transition at $T_{\rm N} \sim$ 1.9~K decreases monotonically up to 1.7~GPa and disappears discontinuously at $P_{\rm c}\sim$ 1.9~GPa, where the structural phase transition at $T_2$ is also suppressed.
At $P_{\rm c}$, enhancement of the effective mass and residual resistivity are observed. 
Another probable magnetic phase transition, however, is found to take place at $T_{\rm ML}\sim$0.3~K and smoothly varies even when crossing $P_{\rm c}$.
Structural phase transitions at ambient pressure were also studied by a single crystal X-ray diffraction measurement.
Below $T_2$, a tetragonal lattice distortion with doubling of the unit cell along the $c$-axis is observed, and thus there are two inequivalent Yb-sites below $P_{\rm c}$.
These results indicate that one Yb-site, having larger valence configuration and thus smaller Kondo temperature, orders at $T_{\rm N}$.
The other phase transition at $T_{\rm ML}$ seems to couple to the crystal structure for $T_2 < T < T_1$. 
YbPd is a unique system exhibiting magnetic orderings in metallic charge-ordered Yb systems.
}

\kword{YbPd, charge ordering, magnetic ordering, pressure-temperature phase diagram}

\begin{document}

\maketitle

\section{Introduction}

Various novel ground states of many Ce and Yb based intermetallic compounds have been considerably studied \cite{Flouquet2005a, Lohneysen2007}.
Unconventional superconductivity and deviation from Fermi liquid behavior due to the enhancement of quantum fluctuations at the quantum critical point have been observed, for example, in CeIn$_3$ and $\beta$-YbAlB$_4$ \cite{Mathur1998, Nakatsuji2008}.    
When the doublet ground state in these Kramers ions is realized, the exchange interaction between the conduction electrons and $f$-electrons seems to determine the ground state.
The Ruderman-Kittel-Kasuya-Yosida interaction stabilizes the magnetic ground state, while the screening of the magnetic moment occurs due to the formation of the Kondo singlet.
For Ce systems, the magnetic ordered state is suppressed under pressure, which has often been compared with Doniach's phase diagram \cite{Settai2007}.
Yb-based compounds with a Yb$^{3+}$ configuration, one 4$f$ hole, are thought to be countersystems to the one 4$f$ electron of trivalent Ce-systems.
In contrast to Ce systems, pressure-induced magnetic phase transitions are often observed in Yb-systems.

The differences between Ce and Yb systems have recently been discussed \cite{Flouquet2011, Flouquet2012}.
Because of the different hierarcchies of the energy scales, such as Kondo temperature, 4$f$-band width, and crystal electric fields, in the real lattice Yb-valence varies widely between the non-magnetic $2+(J=0)$ and the magnetic $3+(J=7/2)$, while Ce valence varies only between $3+(J=5/2)$ and $\sim3.2+$.
In the former case of intermediate valence, the Kondo temperature $T_{\rm K}$ could be small due to the narrower 4$f$-band than Ce.
Thus Yb-compounds are expected to have a magnetic and/or heavy fermion state even in the valence fluctuating regime, something which is only realized in the very nearly trivalent configuration for Ce-systems.

Another intriguing phenomenon is charge ordering; Yb$^{3+}$ and Yb$^{2+}$-configurations lie on the sublattices, for example, in semimetallic Yb$_4$As$_3$ \cite{Ochiai1990}.  
In contrast, for metallic cases hybridization between 4$f$ and conduction electrons may lead to charge orderings of different non-integer configurations.
Being distinguished from the homogeneous intermediate valence systems, such materials are called heterogeneous mixed-valence systems.
These compounds have two or more crystallographically inequivalent sites, and the valences for the respective sites are different. 
Because of their different electronic configurations, magnetic ordering may occur for one of the sublattices.
However, the magnetically ordered state is rarely observed in heterogeneous mixed valence systems.
 
YbPd has been suggested as one such metallic heterogeneous mixed-valence system and shows puzzling four phase transitions at 125, 110, 1.9 and 0.5~K, although it crystallizes in a simple CsCl-structure under ambient conditions \cite{Pott1985, Bonville1986}.
The higher temperature transitions at $T_{1}(T_2)$ = 125~K (110~K) have been detected as symmetric sharp peaks in the temperature dependences of thermal expansion coefficient and specific heat, and a clear thermal hysteresis in the electrical resistivity \cite{Pott1985,Mitsuda2009}.
However, no anomalies in the magnetic susceptibility have been reported, implying first-order structural phase transitions \cite{Landelli1980, Mitsuda2009}.
Its structural distortion has been revealed by Raman scattering experiments; the crystal structure of YbPd loses four-fold symmetry at 4~K, and appears to begin to deform below $T_{1}$=125~K \cite{Hasegawa2011}.
At $T_{\rm N}\sim$ 1.9~K, an antiferromagnetic phase transition has been confirmed via magnetic susceptibility and M$\ddot{\rm o}$ssbauer measurements \cite{Pott1985, Bonville1986}.
In contrast to the other Yb-compounds, in which pressure often stabilizes the magnetic ground state, the pressure suppression of $T_{\rm N}$ in YbPd is manifested by a negative thermal expansion coefficient jump \cite{Pott1985, Tokiwa2011}.
In fact, it has been reported that $T_{\rm N}$ decreases with increasing pressure \cite{Sugishima2010}.
In addition, the susceptibility and the thermal expansion show clear thermal hysteresis around $\sim$0.5~K \cite{Pott1985, Tokiwa2011}.
Despite many experimental works in the literature, the origin of the phase transition has not yet been solved.

Moreover, it has been indicated by M$\ddot{\rm o}$ssbauer studies that the non-magnetic and magnetic Yb-sites coexist in almost equal proportions in the ordered state \cite{Bonville1986}.
The magnetization and magnetic entropy are well reproduced by assuming that half of the Yb-ions have the magnetic Yb$^{3+}$ configuration \cite{Mitsuda2009, Tokiwa2011}.
Therefore, it is proposed that half of the Yb-ions form a heavy fermion antiferromagnet and that the structural phase transitions at $T_{1}$ and $T_{2}$ are associated with ordering of the Yb-valence, i.e., the magnetic state close to Yb$^{3+}$ and the nonmagnetic one having intermediate valence \cite{Bonville1986, Sugishima2010, Tokiwa2011}. 
However, there is no experimental evidence of charge ordering in YbPd, such as the the existence of a detailed pressure-temperature phase diagram including the low temperature region and the observation of crystallographic inequivalent Yb-sites.

In this paper, we have carefully determined the pressure-temperature phase diagram via precise electrical resistivity and ac-calorimetry measurements.
On the basis of these measurements, we discuss the interplay between the valence and magnetic instabilities in YbPd.
The magnetic phase below $T_{\rm N}$ only exists up to $P_{\rm c}\sim$ 1.9~GPa, where the phase transition at $T_{2}$ is also suppressed, while another probable magnetic phase transition at $T_{\rm ML}<T_{\rm N}$ still remains across $P_{\rm c}$.
Tetragonal lattice distortion below $T_2$ is observed via single crystal X-ray diffraction measurements at ambient pressure, indicating the existence of the two inequivalent crystallographic Yb-sites.
Because of the resultant different surroundings of Yb-ions, different characteristic energies, namely $T_{\rm K}$ varying with 4$f$-occupation number, cause one of the Yb-sites to magnetic order at $T_{\rm N}$.
The other magnetic phase transition at $T_{\rm ML}$ seems to couple to the structure for $T_2 < T <T_1$.  
Possible scenarios explaining the low temperature magnetism in this heterogeneous mixed-valence system are discussed. 

\section{Experimental}

\begin{figure}[t]
\begin{center}
\includegraphics[clip,width=6cm]{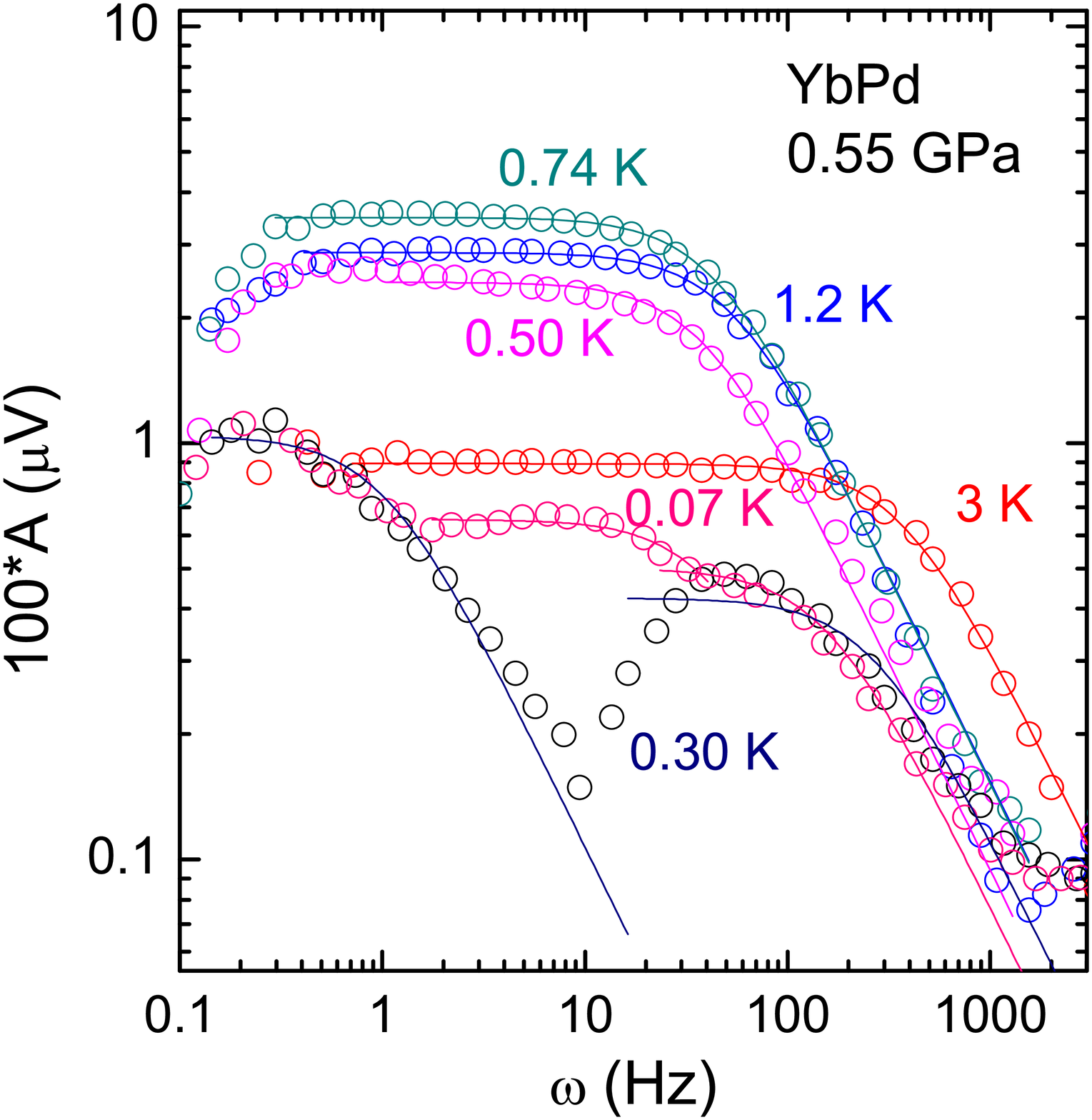}
\end{center}
\caption{(Color online) Frequency dependence of Lock-in-voltage $A$, which is proportional to $T_{\rm ac}$, at 0.55~GPa and several temperatures.
The lines show fitting results for the equation $A\propto T_{\rm ac} = P_0/(\kappa+i \omega C$), where $\kappa$ is the thermal conductivity between sample and the environment.
For ac-calorimetry measurements, $\omega > \kappa/C$ should be satisfied. 
Note that a dispersion $A(\omega)$ appears below $T_{\rm MH}\sim$ 0.5~K, probably reflecting phase separation or the coexistence of higher and lower temperature phases.
}
\label{f-dep}
\end{figure}

Single crystals of YbPd were grown by the self flux method \cite{Canfield1992}.
Typical dimensions of the grown crystals were 1$\times$1$\times$1~mm$^3$.
A CuBe diamond-anvil pressure cell filled with glycerin as a pressure-transmitting medium was used for applying pressure.
The pressure was changed at room temperature and determined by a ruby fluorescence method at 10~K.
The pressure medium remained liquid up to $\sim$5~GPa and at room temperature \cite{Tateiwa2009, Klotz2012}, indicating that hydrostatic conditions were well satisfied in the studied pressure range.
For simultaneous electrical resistivity $\rho$ and ac-calorimetry $C_{\rm ac}$ measurements, two Au-Au$\underbar{\rm Fe}$(0.07\%) thermocouples (TC) and two gold wires were welded on the edges of a rectangular shaped sample with a size of $\sim200~\mu{\rm m}\times 80~\mu{\rm m}\times 40~\mu{\rm m}$.
$\rho$ was measured by the four probe ac method down to 60~mK.
For $C_{\rm ac}$ measurements, the sample temperature was modulated by the ac current through the sample between 60~mK and 4~K.
The suitable modulating frequency was usually determined by the frequency dependence of lock-in-voltage $A$, as shown in Fig.\ref{f-dep}.
At 0.3~K, non trivial dispersion of $A$ around $\omega\sim$10~Hz was seen, and it was suppressed at 0.07~K. 
As a result, the measuring frequency in this work was fixed at $\omega$ = 203~Hz, which is well above the frequency showing anomaly in $A(\omega)$ below 0.5~K and satisfies the measuring condition.
Taking into account the phase shift between heater and TC, $\phi$, we calculate $C_{\rm ac}$ using the formula; $C_{\rm ac} = -P_0\sin\phi/\omega|T_{\rm ac}|$, where $P_0$ is the heating power.
The relative change of $C_{\rm ac}$ as a function of pressure above $T_{\rm N}$ were directly comparable, because the same modulation current and frequency, and the same experimental electronics were used for each pressure. 

Single crystal X-ray diffraction measurements were performed with a four-circle X-ray diffractometer attached to a rotating anode Mo $K_\alpha$ X-ray generator at ambient pressure. 
The sample temperature was controlled using a closed cycle refrigerator.

\section{Experimental results}

\subsection{Pressure dependence of $T_{1}$ and $T_{2}$}

\begin{figure}[t]
\begin{center}
\includegraphics[clip,width=6.5cm]{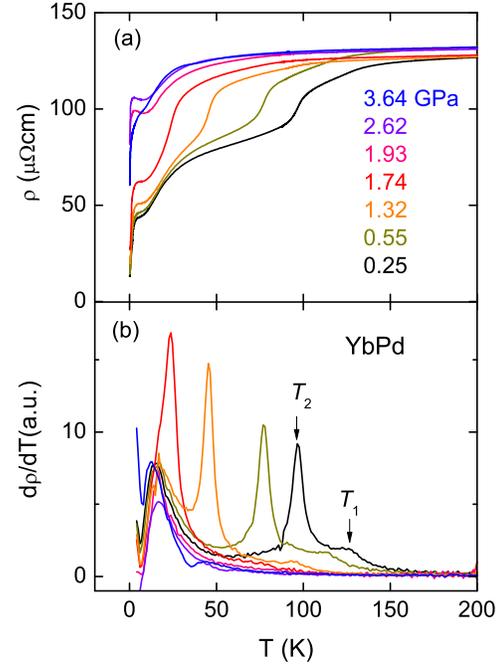}
\end{center}
\caption{(Color online) Temperature dependences of (a)$\rho$ and (b) the temperature derivative ${\rm d}\rho/{\rm d}T$ of YbPd under different pressures.
The arrows indicate the structural transition temperatures $T_{\rm 1}$ and $T_{\rm 2}$.}
\label{rhoT}
\end{figure}

First, we address the pressure dependence of the high temperature structural phase transitions at $T_{1}$ and $T_{2}$ as observed from electrical resistivity measurement. 
Fig.~\ref{rhoT}(a) represents the temperature dependence of the resistivity $\rho(T)$ at several pressures.
In the high temperature region above $T_{1}$, $\rho(T)$ is insensitive to temperature but drops at $T_{1}$ and $T_{2}$ on cooling.
These drops are clearly seen as peaks in the ${\rm d}\rho/{\rm d}T$ curves in Fig.~\ref{rhoT}(b).
$T_{1}$ and $T_{2}$ are observed to decrease with increasing pressure.
This behavior is in agreement with previously reported resistivity and thermal expansion measurements for polycrystalline samples \cite{Mitsuda2009, Sugishima2010}.
A broad peak corresponding to ${\rm d}\rho/{\rm d}T$ of around 16~K is also observed and is insensitive to pressure.
A similar anomaly has been exhibited in the specific heat measurements \cite{Pott1985}, although its origin is not clear.
Above 1.74~GPa, $T_{2}$ is suppressed.
Interestingly, at low temperature $\rho$ is strongly enhanced; this enhancement will be discussed later.
Due to smearing of the ${\rm d}\rho/{\rm d}T$ anomaly at $T_{1}$ with increasing pressure, $T_{1}(P)$ can not be unambiguously determined above 2~GPa.
However, previously reported thermal expansion measurements clearly detected the transition at $T_{1}$ at up to 1.7~GPa, implying it still exists at higher pressure \cite{Sugishima2010}. 
$T_{1}$ may become 0~K at a pressure of~$\sim$4~GPa by linear extrapolation, which agrees with previous report \cite{Sugishima2010}.
These results suggest that the ground state changes while crossing these pressures, accompanied by deformation of the crystal structures for $T < T_2$, $T_2 < T < T_1$ and $T > T_1$, an important finding. 

\subsection{Pressure dependence of magnetic transitions}

\begin{figure}[t]
\begin{center}
\includegraphics[clip,width=6.5cm]{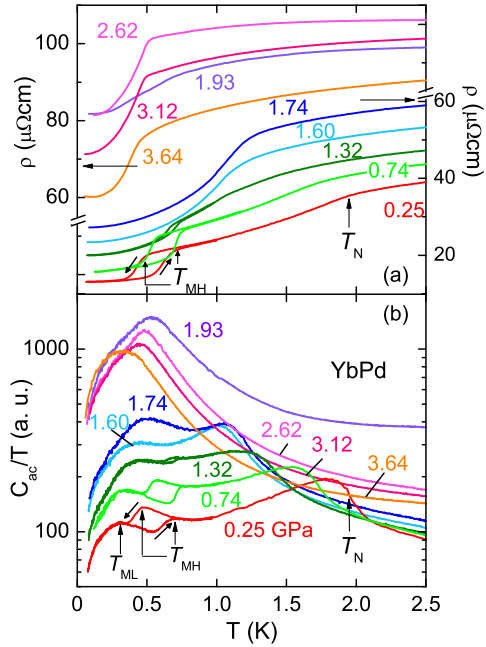}
\end{center}
\caption{(Color online) Temperature dependences of (a) $\rho$ and (b) $C_{\rm ac}/T$ (semi-log plot) of YbPd under different pressures.
The right and left scales in (a) are for below and above 1.74~GPa, respectively.
The arrows indicate $T_{\rm N}$, $T_{\rm MH}$, and $T_{\rm ML}$.
The thermal hysteresis around $T_{\rm MH}$ for $P < P_{\rm cr}\sim$1.45~GPa is also notable.
}
\label{RhoCac}
\end{figure}

Precise simultaneous resistivity and ac-calorimetry measurements revealed the rich magnetic pressure-temperature phase diagram of YbPd. 
Figures \ref{RhoCac} (a) and (b) represent the temperature dependence of the resistivity $\rho$ and the ac specific heat divided by temperature $C_{\rm ac}/T$ under several pressures and below 2.5~K. 
At the lowest measured pressure of $P\sim$~0.25~GPa, a kink in $\rho$ and a $\lambda$-shaped anomaly in $C_{\rm ac}/T$ appear at $T_{\rm N}\sim$1.9~K.
This temperature is in agreement with previous reports \cite{Pott1985, Tokiwa2011}. 
On further cooling, a drop in $\rho$ and a peak in $C_{\rm ac}/T$ are observed at 0.47~K, also as previously reported \cite{Tokiwa2011}. 
This temperature is hereafter labeled $T_{\rm MH}$.
The transition at $T_{\rm MH}$ is obviously first order, due to the observation of thermal hysteresis.
The first order phase transition at $T_{\rm MH}$ may be a transition between two magnetic structures, such as a transition between incommensurate and commensurate.
The dispersion of $A(\omega)$ below 0.5~K in Fig.~\ref{f-dep} may relate to this first order phase transition.
Surprisingly, another peak appears in $C_{\rm ac}/T$ at $\sim$ 0.30~K, while no anomaly in $\rho(T)$ is observed. 
While two similar peaks in the susceptibility at 0.3~K and 0.7~K on warming have previously been reported \cite{Pott1985}, the authors did not declare whether the anomaly corresponded to the phase transition.
We confirm the existence of a phase transition thermodynamically, and the transition temperature is hereafter labeled as $T_{\rm ML}$.
Because of the doublet ground state, the phase transition at $T_{\rm ML}$ is magnetic in origin.
This will be discussed in detail later.
Thus, in addition to the known four phase transitions, we have newly found a magnetic phase transition.

With increasing pressure, $T_{\rm N}$ decreases monotonically, $T_{\rm MH}$ increases with suppression of the hysteresis loop, and $T_{\rm ML}$ weakly increases.
The hysteresis loop around $T_{\rm MH}$ closes at $P\sim$1.45~GPa and $T\sim$0.72~K.
$T_{\rm MH}$ disappears as if the first order transition has changed to a crossover.
Above 1.9~GPa, any anomalies corresponding to the phase transition at $T_{\rm N}$ are not observed, indicating that the transition is suppressed discontinuously.
Moreover, both $\rho(T)$ and $C_{\rm ac}(T)/T$ curves are drastically changed and their values are enhanced, as shown in Fig.~\ref{RhoCac}.
At 1.93~GPa, $C_{\rm ac}(T)/T$ only shows a broad peak at $\sim$ 0.6~K, and a broad kink in $\rho(T)$ is observed correspondingly.
At higher pressure this broad kink in $\rho(T)$ is more pronounced. 
The temperature connects to $T_{\rm ML}$ at lower pressure, implying that the anomalies correspond to the phase transition at $T_{\rm ML}$.
Finally, $T_{\rm ML}$ varies smoothly with pressure, while $T_{\rm N}$ disappears around 1.9~GPa.

\subsection{Pressure-temperature phase diagram}
\begin{figure}
\begin{center}
\includegraphics[clip,width=6.5cm]{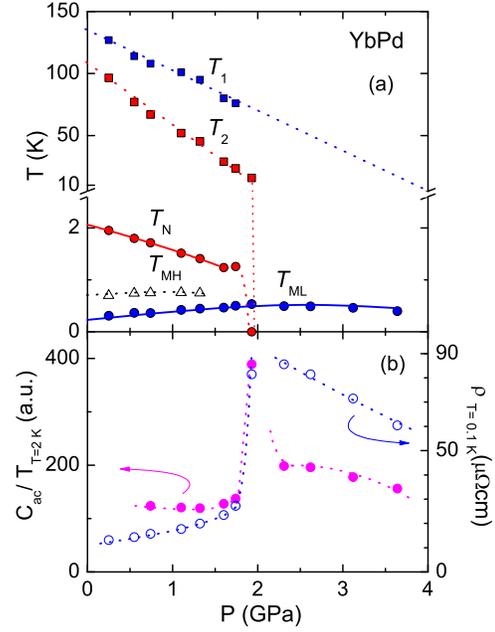}
\end{center}
\caption{(Color Online) (a) Pressure temperature phase diagram of YbPd. 
$T_1$, $T_2$, and $T_{\rm MH}$ on warming are plotted.
(b) Pressure dependence of the $C_{\rm ac}/T$ in the paramagnetic phase of T = 2~K (left scale) and $\rho$ at 0.1~K (right scale).   
}
\label{PTPD}
\end{figure}

The pressure-temperature phase diagram of YbPd is depicted in Fig~\ref{PTPD}~(a). 
With increasing pressure, $T_{1}$, $T_{2}$, and $T_{\rm N}$ decrease, while $T_{\rm MH}$ and $T_{\rm ML}$ increase.
The hysteresis and anomalies in $C_{\rm ac}(T)/T$ and $\rho(T)$ at $T_{\rm MH}$ disappear above 1.5~GPa.
These results suggest that the probable magnetic phase transition changes to a crossover.
Interestingly, the magnetic ordering at $T_{\rm N}$ couples strongly to the structural phase transition at $T_{\rm 2}$.
$T_{\rm N}$ is suppressed discontinuously at $P_{\rm c}\sim$1.9~GPa, where the transition at $T_{2}$ also collapses, as seen in Fig.~\ref{rhoT}.
These results mean that $P_{\rm c}$ is first order in nature.
$T_{\rm N}$ only exists below $T_{2}$.

As shown in Fig.~\ref{PTPD}~(b), the effective mass deduced by $C_{\rm ac}/T$ at $T$ = 2~K in the paramagnetic state shows a sharp peak.
Here, the mass enhancement seems to be driven by the change in crystal structure and the collapse of magnetism.
In the case of CeIn$_3$, mass enhancement is observed at 2.5~GPa \cite{Knebel2001}, where the antiferromagnetism collapses discontinuously \cite{Kawasaki2008}.
Similarly, for YbPd the small energy of $T_{\rm N}\sim 1.2$~K seems to give rise to a significant fluctuation and thus mass enhancement even at the first order critical point.
In addition, $\rho$ at 0.1~K, approximately the residual resistivity, is strongly enhanced across $P_{\rm c}$.
This may be due to the change in crystal structure and the resultant valence.
Theoretically, it is proposed that the enhancement of residual resistivity occurs at the critical point or even crossover of the valence change \cite{Miyake2002}.   
These characteristics suggest that the charge ordering occurs at $T_{\rm 2}$ and is suppressed at $P_{\rm c}$.
At ambient pressure, however, the averaged Yb-valence of YbPd is $\sim 2.8+$ and is insensitive to temperature even when crossing $T_1$ and $T_2$.  
Therefore, the structural phase transitions at $T_{1}$ and/or $T_{2}$ are expected to make the Yb-sites inequivalent. 

\subsection{X-ray diffraction measurements using single crystal}

The cubic structure observed at room temperature changes to a tetragonal structure at low temperature \cite{Bonville1986}.
Moreover, Raman scattering measurements revealed that the symmetry of the crystal structure at 4~K is lower than the tetragonal symmetry \cite{Hasegawa2011}.
To address the structural phase transitions at $T_{\rm 1}$ and $T_{\rm 2}$ in more detail, we performed X-ray diffraction measurements using an as-grown single crystal at ambient pressure. 
Here, the preliminary results explaining the phase diagram are presented,
while the details will be reported elsewhere \cite{Takahashi2012}.

\begin{figure}[t]
\begin{center}
\includegraphics[clip,width=8.5cm]{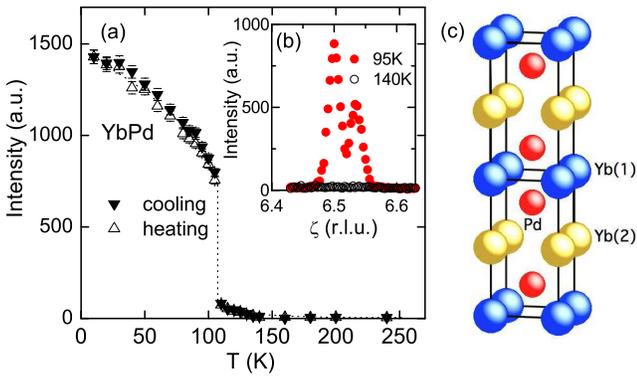}
\end{center}
\caption{(Color online) (a) Temperature dependence of the superlattice reflection intensity at $Q_{\rm s}$ = (0, 0, 6.5) at ambient pressure. 
The full and open symbols indicate data obtained on cooling and heating, respectively.
(b) Superlattice peak profiles along the (0~0~$\zeta$) line measured at 140~K (above $T_1$) and 95~K (below $T_2$).
The peak splitting is due to the Mo $K\alpha_1$ and $K\alpha_2$.    
(c) The most probable structural model for LT-phase ($T<T_2$).
The unit cell is doubled along the $c$-axis due to the displacement of Pd-atoms.
The distance of Yb(1)-Pd is shorter than that of Yb(2)-Pd.}
\label{XRD}
\end{figure}

Splitting of the Bragg reflection at $Q_{\rm B} = (0,~0,~l)$, where $l$ is an integer, was clearly observed below $T_{1}$ (not shown), indicating that the symmetry below $T_{1}$ is lower than cubic.
Moreover, an intense superlattice reflection at $Q_{\rm s} = (0,~0,~6.5)$ was observed below $T_{2}$ as shown in Fig.~\ref{XRD}(b).
The superlattice reflection intensity discontinuously appears at $T_{2}$ and increases on further cooling, confirming a first order nature (Fig.~\ref{XRD}(a)). 
No other superlattice reflections are observed at either $(n/2~n/2~n/2)$ or $(n/2~n/2~0)$.
These results indicate that below $T_2$ the crystal structure is tetragonal with a doubled unit cell along the $c$-axis.
At 10~K, the $c/a$ ratio is approximately 1.007, in good agreement with the previously reported value \cite{Bonville1986}. 
The most probable structure model below $T_2$ is that the Pd atoms displace along the $c$-axis, resulting in the two inequivalent Yb-sites, as shown in Fig.~\ref{XRD}(c).  
The Yb-sites with shorter and longer distances to Pd are labeled as Yb(1) and Yb(2), respectively.
The valence of the Yb(1)-ion is expected to be larger than that of Yb(2), because the radius of the Yb$^{3+}$ ion is smaller than that of Yb$^{2+}$.
Below $P_{\rm c}$, the Yb(1) sublattice seems to order magnetically.

In this study, we could not observe any breaking of the four-fold symmetry as reported from the Raman scattering measurements \cite{Hasegawa2011}.
It is important to reveal whether the four-fold symmetry is preserved at lower temperature.
In any case, two inequivalent Yb-sites exist below $T_{\rm 2}$.  
 
\section{Discussion}
The charge ordering of YbPd has been proposed as the ordering of magnetic and non-magnetic Yb-ions in almost equal quantities at 50~mK \cite{Bonville1986}.  
Here, we have revealed that the structural phase transition from cubic to tetragonal symmetry at $T_{\rm 2}$ causes the Yb-sites to exist as two inequivalent ones, as shown in Fig.~\ref{XRD}(c).
It is expected that the electronic configurations of Yb(1) and Yb(2) are different.
This phase transition is reminiscent of the charge ordering observed in Yb$_4$As$_3$ \cite{Ochiai1990}.
The ordering is also accompanied by the structural phase transition from cubic to trigonal symmetry \cite{Iwasa1998}.
In the charge ordered state, there are two inequivalent Yb-sites; 1/4 of the Yb-ions exist as Yb$^{3+}$ arraed along the $[ 1~1~1]$-direction and the other 3/4 Yb-ions in the unit cell are Yb$^{2+}$ \cite{Staub2001}.
The mean Yb-valence is thus expected to $\sim$2.25+ from the ratio of the divalent and trivalent Yb-ions, which is in good agreement with the effective moment \cite{Ochiai1990}.
In contrast to metallic YbPd, Yb$_4$As$_3$ is a semi-metal having an extremely low carrier concentration of 0.001 per Yb-atom \cite{Ochiai1990}.
A difference between high and low carrier systems appears in their resistive anomaly at charge ordering temperature.
The latter exhibits a sudden increase in the resistivity because the charge ordering decreases the carrier density.
In contrast, for YbPd the resistivity decreases at $T_1$ and $T_2$, as shown in Fig.~\ref{rhoT}(a).
The carrier density is almost constant, because the average valence is insensitive to temperature, and the decrease in resistivity arises from the suppression of valence fluctuation.  

For YbPd, the average valence is $\sim$2.8 and is insensitive to temperature.
Therefore, charge ordering of the divalent and trivalent configurations in equal proportions is ruled out.
It is expected that the charge is screened through the hybridization between conduction and $f$ electrons ($c$-$f$ hybridization), and thus the intermediate valence state may be stabilized in such metallic systems. 
Some Yb compounds, for example Yb$_2$Pt$_3$Sn$_5$\cite{Muro2003} and Yb$_5$Si$_3$\cite{Rams1997}, and YbPd$_3$S$_4$ \cite{Bonville2003}, are known to be heterogeneous mixed-valence systems. 
In stark contrast to the semiconducting Yb$_4$As$_3$, for metallic systems the intermediate valence of the Yb-ions lie on the ordered sublattices.
The interesting and remarkable phenomena exhibited by YbPd compared to these heterogeneous mixed-valence systems are the magnetic orderings in the charge ordered phases, and that they can be tuned by altering the external pressure.
This is due to the moderate balance between charge and spin screenings through the $c$-$f$ hybridization.

The Yb$^{3+}$-ion in the non-cubic lattice splits the $J=7/2$ multiplet into four Kramers doublets for YbPd below $T_2$. 
For the doublet ground state in zero applied field, magnetic ordered or heavy fermion states could be realized to lift the degeneracy in the case for $\Delta_{\rm CEF}>k_{\rm B}T_{\rm K}$, where $\Delta_{\rm CEF}$ is the splitting energy between ground and first excited states \cite{Flouquet2011, Flouquet2012}.
For YbPd, the observation of CEF inelastic scattering indicates $T_{\rm K}$ is comparable to $\Delta_{\rm CEF}
$ \cite{Walter1987}.
It is also known that $T_{\rm K}$ varies with the occupation number of the 4$f$ level $n_f$, 
\begin{equation}
k_{\rm B}T_{\rm K}\sim\frac{(1-n_{f})}{n_f}\Delta_{4f}N_f,
\label{TK}
\end{equation}
where $\Delta_{4f}$ and $N_f$ are the 4$f$-band width and the degeneracy of the 4$f$ multiplet, respectively \cite{Rice1986, Malterre1996, Flouquet2005}. 
Owing to the much smaller $\Delta_{4f}$ of Yb than Ce, $T_{\rm K}$ could be smaller or comparable to $\Delta_{\rm CEF}$ even in the intermediate-valence state ($n_f\ll1$), which is in strong contrast to the Ce-case.
This leads to intermediate valence Yb-systems in the magnetic ordering or paramagnetic heavy fermion states, which are usually restricted to the trivalent ($n_f =1$) configuration in Ce.
 
The tetragonal distortion at $T_{\rm 2}$ excludes the proposed $\Gamma_8$ quartet ground state, and thus the possibility of quadrupolar ordering \cite{Pott1985, Walter1987}. 
Therefore, magnetic ordering is most probable at both $T_{\rm N}$ and $T_{\rm ML}$.
Yb(1), which has the shorter distance to Pd, is expected to have the larger valence due to the smaller ionic radius of Yb$^{3+}$ than that of Yb$^{2+}$, and the resultant smaller $T_{\rm K}$ expected from Eq.~(\ref{TK}).
Conversely, Yb(2) has larger $T_{\rm K}$.
The different $T_{\rm K}$s seem to give rise to the different magnetic ordering transition temperatures for Yb(1)- and Yb(2)-sublattices, i.e. $T_{\rm N}$ and $T_{\rm ML}$. 
From the above consideration, both transitions would be expected to disappear at $P_{\rm c}$, where $T_{2}$ is suppressed.
In fact, while $T_{\rm N}$ only exists below $T_{2}$, the phase transition at $T_{\rm ML}$ still remains above $P_{\rm c}$.
Thus, the $T_{\rm ML}$ transition seems to couple to the $T_{1}$ phase transition.
These results indicate that charge orderings near and far from the trivalent configurations occur not only at $T_{2}$ but also at $T_{1}$.

While the existence of the phase transition at $T_{\rm ML}$ has been clearly observed both from the present ac-calorimetry results and previous susceptibility studies \cite{Pott1985}, it could not be detected at ambient pressure in some experiments. 
The hyperfine field is not significantly changed at either $T_{\rm MH}$ or $T_{\rm ML}$ \cite{Bonville1986}.
The resistivity also shows no anomaly at $T_{\rm ML}$ below $P_{\rm c}$, as seen in Fig.\ref{RhoCac}(b).
The change in the magnetic scattering, probably in proportion to the change in the ordered moment, is very small.    
This may be in agreement with the lack of an anomaly in the hyperfine field; the experimental resolution for $^{140}$Yb isotope M$\ddot{\rm o}$ssbauer has been reported to be 0.15~$\mu_{\rm B}$ \cite{Bonville1994}.
It may also be in agreement with the low transition temperature of $T_{\rm ML}\sim$0.3~K.
The ordered moment is significantly reduced through the $c$-$f$ hybridization.
In addition, the specific heat and thermal expansion results measured for a polycrystalline sample exhibited anomalies corresponding to the phase transition at $T_{\rm N}$ and $T_{\rm MH}$, but not at $T_{\rm ML}$ \cite{Tokiwa2011}. 
Because the phase transition at $T_{\rm ML}$ couples with the crystal structure for $T_2 < T < T_1$, the polycrystalline crystal averages subtle sublattices, and thus the magnetic ordering at $T_{\rm ML}$ is smeared out. 
In addition, below $P_{\rm c}$, the magnetic phase transition at $T_{\rm N}$ seems to smear the transition at $T_{\rm ML}$.
Across $P_{\rm c}$, the anomalies in $\rho(T)$ and $C_{\rm ac}(T)/T$ that correspond to the transition are significantly emphasized.
The clear anomaly in $\rho(T)$ above $P_{\rm c}$ is due to the release of the large residual entropy.
The existence of an interplay between $T_1$ and $T_{\rm ML}$ is further speculated from these results.
We have also observed splitting of the Bragg peaks (not shown) and a small but significant development of the superlattice reflection below $T_1$ (Fig.~\ref{XRD}(a)).
Similarly to the structure observed below $T_2$, some of the Yb-ions have near trivalent configurations in the structure below $T_1$, and the sublattice is expected to order below $T_{\rm ML}$.

At $P_{\rm c}$, the effective mass ($C_{\rm ac}/T$) and the residual resistivity are strongly enhanced.
After peaking, the mass still remains at a larger value than that below $P_{\rm c}$.
This indicates that the magnetically ordered 4$f$ electrons form heavy fermion states through hybridization with the conduction electrons. 
The enhancement of the residual resistivity at $P_{\rm c}$ indicates that the structure and the resultant valence change across $P_{\rm c}$ \cite{Miyake2002}.
A similar enhancement in $\rho_0(P)$ at $P_{\rm c}$ has also been observed in YbInCu$_4$, which shows isostructural phase transition accompanying the first order valence transition at ambient pressure \cite{Park2006}.
These results support the proposed scenario for the charge ordering of YbPd at $T_1$ and $T_2$; trivalent and the intermediate valence states lie on the sublattices. 
It is important to solve the crystal structure for $T_2 < T <T_1$, which is currently in progress \cite{Takahashi2012}.
It is also very important in revealing whether the ordered phase collapses at the pressure suppressing $T_{1}$.

Pressure often stabilizes the magnetic ground state in Yb-systems because Yb$^{3+}$ has a smaller ionic radius than non-magnetic Yb$^{2+}$.
In contrast, for YbPd the magnetic phases are suppressed with pressure as for the Ce Kondo lattices.  
The pressure dependence of Kondo temperature $T_{\rm K}$ and the antiferromagnetic exchange $J$ for Ce- and Yb-based Kondo lattices have been discussed theoretically \cite{Goltsev2005}.  
With increasing pressure, both $T_{\rm K}$ and $J$ show a broad minimum in the case of Yb, while for Ce they increase monotonously.  
YbPd may locate near the pressure of the broad minimum in $T_{\rm K}$ and $J$ in terms of Ref.~\citen{Goltsev2005}.
Moreover, the antiferromagnetic phase is destabilized with suppression of the low temperature phase.
The charge ordering thus plays a very important role in the low temperature magnetism.

It is also interesting how the ground state varies above the expected critical pressure, where $T_1$ collapses.
Above this pressure, the intermediate valence state is stabilized over the whole temperature range in a similar case to that of homogeneous valence fluctuating systems.
The latter materials, for example YbCu$_2$Si$_2$, change ground state from non-magnetic to magnetic at relatively high pressure \cite{Alami-Yadri1998}.
However,  no pressure-induced phase transitions have been observed for YbPd at up to 8~GPa, \cite{Sugishima2010}. 

\section{Conclusion}

Precise electrical resistivity and ac-calorimetry measurements have revealed the unique pressure-temperature phase diagram of YbPd.
In addition to the reported four phase transitions, we have found a lower temperature magnetic phase transition at $T_{\rm ML}\sim$0.3~K. 
Single crystal X-ray diffraction measurements confirmed that the structural phase transition at $T_{2}$ doubles the unit cell along the $c$-axis, resulting in two inequivalent Yb-sites in the tetragonal symmetry.
These results are decisive evidence of charge ordering in the metallic systems.
One Yb-site having larger valence and smaller Kondo temperature may order magnetically at $T_{\rm N}$.
This scenario agrees with the fact that $T_{\rm N}$ is discontinuously suppressed at $P_{\rm c}$, where $T_2$ collapses.
In addition, strong enhancements of the residual resistivity and effective mass are observed at $P_{\rm c}$, suggesting the development of spin and valence fluctuations. 
YbPd is a unique system which shows magnetic ordering in charge ordered states.

\section*{Acknowledgements}
We thank Prof. H. Harima and Prof. K. Miyake for valuable discussions.
This work was supported by the Osaka University Global COE program ``Core Research and Engineering of the Advanced Materials-Interdisciplinary Education Center for Materials Science'' and a JSPS Grant-in-Aid for Young Scientists (B) (No. 23740271), and also by the Exciting Leading-
Edge Research Project ``Heavy Electrons and Superconductivity produced by High Quality Single
Crystals and High Pressure Techniques'' at Osaka University.
TE thanks The Japan Securities Scholarship Foundation.


\begin{thebibliography}{99}

\bibitem{Flouquet2005a} J. Flouquet: in {\it Progress in Low Temperature Physics}, ed. W. P. Halperin (Elsevier, Amsterdam, 2005) Vol. 15, p.139. 

\bibitem{Lohneysen2007} H. v. L$\bar{\rm o}$hneysen, A. Rosch, M. Vojta and P. W$\bar{\rm o}$lfle: Rev. Mod. Phys. {\bf 79} (2007) 1015.
 
\bibitem{Mathur1998} N. D. Mathur, F. M. Grosche, S. R. Julian, I. R. Walker, D. M. Freye, R. K. W. Haselwimmer and  G. G. Lonzarich: Nature {\bf 394} (1998) 39.

\bibitem{Nakatsuji2008} S. Nakatsuji, K. Kuga, Y. Machida, T. Tayama, T. Sakakibara, Y. Karaki, H. Ishimoto, S. Yonezawa, Y. Maeno, E. Pearson, G. G. Lonzarich, L. Balicas, H. Lee and Z. Fisk: Nature Phys. {\bf 4} (2008) 603.

\bibitem{Settai2007} R. Settai, T. Takeuchi and Y. $\bar{\rm O}$nuki: J. Phys. Soc. Jpn. {\bf 76} (2007) 051003. 

\bibitem{Flouquet2011} J. Flouquet and H. Harima: arXiv:0910.3110. 

\bibitem{Flouquet2012} J. Flouquet and H. Harima: {\it Kotai Butsuri} (Solid State Physics) \textbf{47} (2012) 47 [in Japanese].

\bibitem{Ochiai1990} A. Ochiai, T. Suzuki and T. Kasuya: J. Phys. Soc. Jpn. 59 (1990) 4129.

\bibitem{Pott1985} R. Pott, W. Boksch, G. Leson, B. Politt, H. Schmidt, A. Freimuth, K. Keulerz, J. Langen, G. Neumann, F. Oster, J. R$\ddot{\rm o}$hler, U. Walter, P. Weidner and D. Wohlleben: Phys. Rev. Lett. {\bf 54} (1985) 481.

\bibitem{Bonville1986} P. Bonville, J. Hammann, J. A. Hodges, P. Imbert and G. J. J$\acute{\rm e}$hanno: Phys. Rev. Lett. {\bf 57} (1986) 2733.

\bibitem{Landelli1980} A. Landelli, G. L. Olcese and A. Palenzona: J. Less-Cmmon Met. {\bf 76} (1980) 317. 

\bibitem{Mitsuda2009} A. Mitsuda, K. Yamada, M. Sugishima and H. Wada, Physica B {\bf 404} (2009) 3002.

\bibitem{Hasegawa2011} T. Hasegawa, N. Ogita, M. Sugishima, A. Mitsuda, H. Wada and M. Udagawa, J. Phys.: Conf. Ser. {\bf 273} (2011) 012030.

\bibitem{Tokiwa2011} Y. Tokiwa, S. Gr$\ddot{\rm u}$nheit, H. S. Jeevan, C. Stingl and P. Gegenwart: J. Phys.: Conf. Ser. {\bf   273} (2011) 012062.

\bibitem{Sugishima2010} M. Sugishima, K. Yamada, A. Mitsuda, H. Wada, K. Matsubayashi, Y. Uwatoko, K. Suga and K. Kindo: J. Phys.: Condens. Matter {\bf 22} (2010) 375601.

\bibitem{Canfield1992} P. C. Canfield and Z. Fisk: Phil. Mag. B {\bf 65} (1992) 1117.

\bibitem{Tateiwa2009} N. Tateiwa and Y. Haga: Rev. Sci. Instrum {\bf 80} (2009) 123901.

\bibitem{Klotz2012} S. Klotz, K. Takemura, Th. Str$\ddot{\rm a}$ssle and Th. Hansen: J. Phys.: Condens. Matter {\bf 24} (2012) 325103.
 
\bibitem{Knebel2001} G. Knebel, D. Braithwaite, P. C. Canfield, G. Lapertot and J. Flouquet: Phys. Rev. B {\bf 65} (2001) 024425.

\bibitem{Kawasaki2008} S. Kawasaki, M. Yashima, Y. Kitaoka, K. Takeda, K. Shimizu, Y. Oishi, M. Takata, T. C. Kobayashi, H. Harima, S. Araki, H. Shishido, R. Settai and Y. $\bar{\rm O}$nuki: Phys. Rev. B {\bf 77} (2008) 064508.

\bibitem{Miyake2002} K. Miyake and H. Maebashi: J. Phys. Soc. Jpn. {\bf 71} (2002) 1007.

\bibitem{Takahashi2012} R. Takahashi, T. Honda, A. Miyake, T. Kagayama, K. Shimizu, T. Ebihara, T. Kimura and Y. Wakabayashi: arXiv:1303.4831.

\bibitem{Walter1987} U. Walter and D. Wohlleben: Phys. Rev. B {\bf 35} (1987) 3576.

\bibitem{Iwasa1998} K. Iwasa, M. Kohgi, N. Nakajima, R. Yoshitake Y. Hisazaki, H. Osumi, K. Tajima, N. Wakabayashi, Y. Haga, A. Ochiai, T. Suzuki and A. Uesawa: J. Magn. Magn. Mater. {\bf 177-181} (1998) 393.

\bibitem{Staub2001} U. Staub, B. D. Patterson, C. Schulze-Briese, F. Fauth, M. Shi, L. Soderholm, G. B. M. Vaughan and A. Ochiai: Europhys. Lett., {\bf 53} (2001) 72.

\bibitem{Muro2003} Y. Muro, K. Yamane, M.-S. Kim, T. Takanatake, C. Godart and P. Rogl: J. Phys. Soc. Jpn. {\bf 72}(2003) 1745.

\bibitem{Rams1997} M. Rams, K. Kr$\acute{\rm o}$las, P. Bonville, E. Alleno, C. Godart, D. Kaczorowski and P. Canepa: Phys. Rev. B {\bf 56} (1997) 3690.

\bibitem{Bonville2003} P. Bonville, C. Godart, E. Alleno, F. Takahashi, E. Matsuoka and M. Ishikawa: J. Phys.: Condens. Matter {\bf 15} (2003) L263.

\bibitem{Rice1986} T. M. Rice and K. Ueda: Phys. Rev. B {\bf 34} (1986) 6420.

\bibitem{Malterre1996} D. Malterre, M. Grioni and Y. Baer: Adv. Phys. {\bf 45} (1996) 299.

\bibitem{Flouquet2005} J. Flouquet, A. Barla, R. Boursier, J. Derr and G. Knebel: J. Phys. Soc. Jpn. {\bf 74} (2005) 178.

\bibitem{Bonville1994} P. Bonville, A. Ochiai, T. Suzuki and E. Vincent: J. Phys. I France {\bf 4} (1994) 595.

\bibitem{Park2006} T. Park, V. A. Sidorov, J. L. Sarrao and J. D. Thompson: Phys. Rev. Lett. {\bf 96} (2006) 046405.

\bibitem{Goltsev2005} A. V. Goltsev and M. M. Abd-Elmeguid: J. Phys.: Condens. Matter {\bf 17} (2005) S813. 

\bibitem{Alami-Yadri1998} K. Alami-Yadri, H. Wilhelm and D. Jaccard: Eur. Phys. J. B {\bf 6} (1998) 5.

\end{thebibliography}
\end{document}